\documentstyle[pra,aps,multicol]{revtex}

\begin{document}

\title{{\huge \bf Some aspects of separability}}

\author{{\large Shengjun Wu and Jeeva Anandan}}

\address{{\it Department of Physics and Astronomy, University of South
Carolina, Columbia, South Carolina, SC 29208, USA}}

\date{Oct. 31, 2001, revised Jan. 8, 2002}

\maketitle

\vskip 0.2in

\begin{abstract}
\large
We present three necessary separability criteria for
bipartite mixed states,  the violation of each of these conditions is
a sufficient condition for entanglement.
Some ideas on the issue of finding a necessary and
sufficient criterion of separability are also discussed.
\end{abstract}

\vskip 0.3in

\large

A bipartite quantum state $\rho_{AB}$ is called separable (disentangled) iff
it can be written as 
\begin{equation}
\rho_{AB}=\sum_i q_i \rho_i^A \otimes \rho_i^B
\end{equation}
where $\rho_i^A$ and $\rho_i^B$ are density matrices for systems A and B
respectively, and $q_i > 0$, $\sum_i q_i =1$.  This condition is
equivalent to the condition 
\begin{equation}
\rho_{AB}=\sum_i p_i \left| \psi_i \right\rangle \left\langle \psi_i \right|
\otimes \left| \phi_i \right\rangle\left\langle \phi_i \right|
\end{equation}
where $ p_i > 0 $, $\sum_i p_i =1$, and $\left| \psi_i
\right\rangle$ , $\left| \phi_i \right\rangle$ are normalized pure states of
systems A and B respectively, they may not be orthogonal in general.

So far, there have been many ingenious separability criteria. Since a
separable state always satisfies Bell's inequalities, the latter represent a
necessary condition for separability \cite{Be,We}, but generally they are
not sufficient. Peres \cite{Pe} discovered another simple necessary
condition for separability, a partial transposition of a bipartite quantum
state $\rho _{AB}$ with respect to a subsystem $A$ (or $B$) must be positive
if $\rho _{AB}$ is separable. Peres' criterion has been shown by Horodecki
et al. \cite{MPR} to be strong enough to guarantee separability for
bipartite systems of dimension $2\times 2$ or $2\times 3$, but, for other
cases it is not a sufficient one. It has been proved by Horodecki et al. 
\cite{MPR} that a necessary and sufficient condition for separability of
bipartite mixed state is its positivity under all the maps of the form $%
I\otimes \Lambda$, where $\Lambda$ is any positive map. This criterion is
more important in theory than in practice since it involves the
characterization of the set of all positive maps, which is not easy.
Horodecki-Horodecki\cite{MP} and Cerf-Adami-Gingrich \cite{cag} have
independently derived a reduction criterion of separability for bipartite
quantum states. This criterion is equivalent to Peres' for $2\times n$
composite systems, and it is not sufficient for separability in general
cases. Many interesting separability criteria have been presented recently,
such as the rank separability criterion derived by Horodecki et al. \cite
{hstt}, according to which a separable state cannot have the rank of a reduced
density matrix greater than the rank of total density matrix. This necessary
condition is easy to apply. It is shown that the separability criterion
is also equivalent to solving a set of equations \cite{Wu}; for a given
quantum state which has few nonzero eigenvalues, an analytic result is
possible, but for general states, only numerical approach is practical. More
recently, Nielsen et al \cite{Nielsen} presented the disorder criterion for
separability: a separable state has the property that the vector of
eigenvalues of the density matrix of system AB is majorized by the vector of
eigenvalues of the density matrix of system A (B) alone.

In the following, we shall present three necessary separability criteria for
bipartite mixed states; and a state which violates one of these criteria must be
entangled. Similar separability criteria can be obtained for multipartite
mixed states. Some ideas on the issue of finding a necessary and sufficient
criterion for separability will also be discussed.

{\bf Theorem 1:} If $\rho_{AB}$ is separable then 
\begin{equation}
\left\{
\begin{array}{l}
tr_{A} (\rho_{A}^2) \ge tr_{AB} (\rho_{AB}^2) \\ 
tr_{B} (\rho_{B}^2) \ge tr_{AB} (\rho_{AB}^2)
\end{array}
\right.  \label{theorem1}
\end{equation}
where $\rho_A$, $\rho_B$ are the reduced density matrices of systems A and B
respectively.

Proof. Since $\rho_{AB}$ is separable, i.e., 
\[
\rho_{AB} =\sum_i p_i \left| \psi_i^A \phi_i^B \right\rangle \left\langle
\psi_i^A \phi_i^B \right| 
\]
we have 
\[
\begin{array}{l}
\rho_{A} =\sum_i p_i \left| \psi_i^A \right\rangle \left\langle \psi_i^A
\right| \\ 
\rho_{B} =\sum_i p_i \left| \phi_i^B \right\rangle \left\langle \phi_i^B
\right|
\end{array}
\]
Therefore 
\begin{eqnarray*}
tr_A (\rho_{A}^2) & = & \sum_{ii\prime} p_i p_{i\prime} \left| \left\langle
\psi_i^A | \psi_{i\prime}^A \right\rangle\right|^2 \\
&\ge & \sum_{ii\prime} p_i p_{i\prime} \left| \left\langle \psi_i^A \phi_i^B
| \psi_{i\prime}^A \phi_{i\prime}^B \right\rangle\right|^2 \\
& = & tr_{AB} (\rho_{AB}^2)
\end{eqnarray*}
The same argument holds for the second inequality.

Remark: (i) This criterion is very easy to operate, we even need not
calculate the eigenvalues. The drawback is that although it's a sufficient
separability criterion for pure states, it is not a sufficient condition for mixed
states, not even for the 2-qubit cases. (ii) This condition can also
obtained from Nielsen et al's disorder criterion \cite{Nielsen}, so it is
not as strong as the disorder criterion. (iii) This criterion can be
extended easily. A tripartite separable state 
\begin{equation}
\rho_{ABC}=\sum_i p_i \left| \psi_i^A \right\rangle \left\langle \psi_i^A
\right| \otimes \left| \phi_i^B \right\rangle \left\langle \phi_i^B \right|
\otimes \left| \varphi_i^C \right\rangle \left\langle \varphi_i^C \right|
\end{equation}
always satisfies the following conditions 
\begin{equation}
tr_{\alpha} \{ \rho_{\alpha}^2 \} \ge tr_{\alpha \beta} \{ \rho_{\alpha
\beta}^2 \} \ge tr_{ABC} \{ \rho_{ABC}^2 \}
\end{equation}
where $\alpha$, $\beta$ $\in \{ A, B, C \} $, $\alpha\neq \beta$, and $%
\rho_{\alpha}$ $(\rho_{\alpha \beta})$ is the reduced density matrix of
system $\alpha$ (systems $\alpha + \beta$).

This theorem is generalized to a $n-$partite separable state as follows. 
Let $\rho_{A_1 A_2 ...A_n}$ be the density matrix of $n$ systems
$A_1,A_2,...A_n$. 
If this state 
is separable then it may be written as 
\begin{equation}
\rho_{A_1 A_2 ...A_n}=\sum_i p_i |\psi_i^{A_1}\psi_i^{A_2}....\psi_i^{A_n}>
<\psi_i^{A_1}\psi_i^{A_2}....\psi_i^{A_n}|
\end{equation}
It follows that
\begin{equation}
tr_{\alpha_1} \{\rho_{\alpha_1}^2\} \ge tr_{\alpha_1\alpha_2}\{
\rho_{\alpha_1\alpha_2}^2\} 
\geq.....
\geq tr_{\alpha_1 ...\alpha_r}\{\rho_{\alpha_1 ...\alpha_r}^2\}
\geq .....
\geq tr_{A_1 A_2 ...A_n} \{ \rho_{A_1 A_2 ...A_n}^2\}
\label{inequality}
\end{equation}
where $\alpha_1\alpha_2,...$ represent distinct elements of the set  
$\{A_1, A_2,...A_n\}$. Also, $\rho_{\alpha_1}$ is the reduced 
density matrix of $\alpha_1$, $\rho_{\alpha_1\alpha_2}$ is the reduced 
density matrix of $\alpha_1 + \alpha_2$ etc. (i.e. $\rho_{\alpha_1}$
obtained  by tracing over all systems except $\alpha_1$, and
$\rho_{\alpha_1\alpha_2}$
is obtained  by tracing over all systems except $\alpha_1$ and $\alpha_2$ etc.
from $\rho_{A_1 A_2 ...A_n}$).
The theorem (\ref{inequality}) is proved by noting that 
for every integer $r$ satisfying $1\leq r < n$,
$$\rho_{\alpha_1 ...\alpha_r}=\sum_i p_i
|\psi_i^{\alpha_1}....\psi_i^{\alpha_r}>
<\psi_i^{\alpha_1}....\psi_i^{\alpha_r}|$$
Therefore,
\begin{eqnarray*}
tr_{\alpha_1 ...\alpha_r}\{\rho_{\alpha_1 ...\alpha_r}^2\}
&=& \sum_{ii'} p_i p_{i'} |<\psi_i^{\alpha_1}....\psi_i^{\alpha_r}|
\psi_{i'}^{\alpha_1}....\psi_{i'}^{\alpha_r}>|^2     \\
&\geq& \sum_{ii'} p_i p_{i'}
|<\psi_i^{\alpha_1}....\psi_i^{\alpha_r}\psi_i^{\alpha_{r+1}}|
\psi_{i'}^{\alpha_1}....\psi_{i'}^{\alpha_r}\psi_{i'}^{\alpha_{r+1}}>|^2 \\
&=& tr_{\alpha_1 ...\alpha_r\alpha_{r+1}}\{\rho_{\alpha_1
...\alpha_r\alpha_{r+1}}^2\}
\end{eqnarray*}

Actually, as was pointed out in \cite{Nielsen}, attempts to
characterize separability based only upon the eigenvalue spectra
$\lambda(\rho_{AB})$, $\lambda(\rho_A)$, $\lambda(\rho_B)$ cannot work.
The following two states have exactly the same eigenvalue spectra both
locally and globally, 
\begin{eqnarray}
\rho_{AB} & = & \frac{1}{3} \left| 00 \right\rangle \left\langle 00 \right|
+ \frac{2}{3} \left| \psi^+ \right\rangle \left\langle \psi^+ \right| \\
\sigma_{AB} &=& \frac{2}{3} \left| 00 \right\rangle \left\langle 00 \right|
+ \frac{1}{3} \left| 11 \right\rangle \left\langle 11 \right|
\end{eqnarray}
where $\left| \psi^+ \right\rangle \equiv \frac{1}{\sqrt{2}} \left( \left|
01 \right\rangle + \left| 10 \right\rangle \right)$, but $\rho_{AB}$ is
entangled while $\sigma_{AB}$ is separable.

In the following other two necessary separability conditions are presented
for bipartite systems of dimension $2\times N$.
Let the dimension of system A be $2$ and the dimension of system B be $N$. A
given state $\rho _{AB}$ 
\begin{equation}
\rho _{AB}=\left( 
\begin{array}{ll}
\rho _{00} & \rho _{01} \\ 
\rho _{10} & \rho _{11}
\end{array}
\right)   \label{rho}
\end{equation}
of systems A and B, where each $\rho _{kl}$($k,l=0,1$) is
an $N$-dimensional matrix, can also be written as
\begin{equation}
\rho _{AB}=\left( 
\begin{array}{ll}
M_0+M_z & M_x-iM_y \\ 
M_x+iM_y & M_0-M_z
\end{array}
\right)   \label{rhom}
\end{equation}
or 
\begin{equation}
\rho _{AB}=I\otimes M_0+\sigma _x\otimes M_x+\sigma _y\otimes M_y+\sigma
_z\otimes M_z  \label{defms}
\end{equation}
where the four matrices 
\begin{equation}
\left. 
\begin{array}{c}
M_0=\frac 12\left( \rho _{00}+\rho _{11}\right)  \\ 
M_z=\frac 12\left( \rho _{00}-\rho _{11}\right)  \\ 
M_x=\frac 12\left( \rho _{01}+\rho _{10}\right)  \\ 
M_y=\frac i2\left( \rho _{01}-\rho _{10}\right) 
\end{array}
\right.   \label{msrho}
\end{equation}
are $N$-dimensional and Hermitian.

Let $R$ be a $3$-dimensional real matrix, and $\Upsilon _R$ be a
transformation on the density matrix $\rho _{AB}$%
\begin{equation}
\Upsilon _R\left( \rho _{AB}\right) \equiv \left( 
\begin{array}{ll}
M_0+M_z^R & M_x^R-iM_y^R \\ 
M_x^R+iM_y^R & M_0-M_z^R
\end{array}
\right)  \label{transd}
\end{equation}
where 
\begin{equation}
\left( 
\begin{array}{l}
M_x^R \\ 
M_y^R \\ 
M_z^R
\end{array}
\right) =R\left( 
\begin{array}{l}
M_x \\ 
M_y \\ 
M_z
\end{array}
\right)  \label{transm}
\end{equation}

We have the following two theorems.

{\bf Theorem 2}: If $\rho _{AB}$ is separable, then $\Upsilon _R\left( \rho
_{AB}\right) $ must be positively defined (actually it must be a density
matrix) for any $3$-dimensional real matrix $R$ which satisfies $I-R^TR\geq
0 $. (i.e., $I-R^TR$ is positively defined, or equivalently, all eigenvalues
of $R^TR$ are less than or equal to $1$.)

Remark: If we choose 
\[
R=\left( 
\begin{array}{lll}
1 & 0 & 0 \\ 
0 & -1 & 0 \\ 
0 & 0 & 1
\end{array}
\right) 
\]
$\Upsilon _R$ is the partial transpose of system A, because the latter
transformation interchanges $\rho_{01}$ and $\rho_{10}$, therefore we
obtain Peres' criterion as a special case of our theorem 2 for bipartite
systems of dimension $2\times N$. In view of a result of \cite{MPR} mentioned
above, theorem 2 provides a necessary and sufficient
criterion for bipartite systems of dimension $2\times 2$ or $2 \times 3$.

{\bf Theorem 3}: If $\rho _{AB}$ is separable, then

(i) $M_0-\overrightarrow{r}\cdot \overrightarrow{M}$ (i.e., $%
M_0-xM_x-yM_y-zM_z$ ) is positively defined for any vector $\overrightarrow{r%
}$ of unit length or less ($x^2+y^2+z^2\leq 1$); and

(ii) $tr\left( M_0^2-M_x^2-M_y^2-M_z^2\right) \geq 0$.

$M_0$, $M_x$, $M_y$, $M_z$ are defined in eqs. (\ref{rho}-\ref{msrho}).

Proof: A separable bipartite state 
\[
\rho _{AB}=\sum_ip_i\left| \psi _i^A\right\rangle \left\langle \psi
_i^A\right| \otimes \left| \phi _i^B\right\rangle \left\langle \phi
_i^B\right| 
\]
of dimension $2\times N$ can be rewritten as 
\begin{equation}
\left. 
\begin{array}{lll}
\rho _{AB} & = & \sum_ip_i\frac 12\left( I+\overrightarrow{r_i}\cdot 
\overrightarrow{\sigma }\right) \otimes \left| \phi _i^B\right\rangle
\left\langle \phi _i^B\right|  \\ 
& = & \frac 12\left[ I\otimes \sum_ip_i\left| \phi _i^B\right\rangle
\left\langle \phi _i^B\right| +\sigma _x\otimes \sum_ix_ip_i\left| \phi
_i^B\right\rangle \left\langle \phi _i^B\right| +\sigma _y\otimes
\sum_iy_ip_i\left| \phi _i^B\right\rangle \left\langle \phi _i^B\right|
+\sigma _z\otimes \sum_iz_ip_i\left| \phi _i^B\right\rangle \left\langle
\phi _i^B\right| \right] 
\end{array}
\right.   \label{septh}
\end{equation}
where $\sigma _x$, $\sigma _y$, $\sigma _z$ are the Pauli matrices and \{$%
\overrightarrow{r_i}$\}are real vectors on the bloch spree (i.e., the
coordinators $x_i$, $y_i$, $z_i$ of each vector $\overrightarrow{r_i}$
satisfies $x_i^2+y_i^2+z_i^2=1$). Comparing eqs.(\ref{defms}) and (\ref
{septh}), we have 
\begin{equation}
\left. 
\begin{array}{lll}
M_0 & =  & \frac 12\sum_ip_i\left| \phi _i^B\right\rangle \left\langle
\phi _i^B\right| =\frac 12\rho _B \\ 
M_x & =  & \frac 12\sum_ix_ip_i\left| \phi _i^B\right\rangle
\left\langle \phi _i^B\right|  \\ 
M_y & =  & \frac 12\sum_iy_ip_i\left| \phi _i^B\right\rangle
\left\langle \phi _i^B\right|  \\ 
M_z & =  & \frac 12\sum_iz_ip_i\left| \phi _i^B\right\rangle
\left\langle \phi _i^B\right| 
\end{array}
\right.   \label{sepms}
\end{equation}
From eqs. (\ref{transd}, \ref{transm}, \ref{septh}, \ref{sepms}), we obtain 
\[
\left. 
\begin{array}{lll}
\Upsilon _R\left( \rho _{AB}\right)  & = & \sum_ip_i\frac 12\left( I+%
\overrightarrow{r_i^{\prime }}\cdot \overrightarrow{\sigma }\right) \otimes
\left| \phi _i^B\right\rangle \left\langle \phi _i^B\right|  \\ 
& = & \frac 12\left[ I\otimes \sum_ip_i\left| \phi _i^B\right\rangle
\left\langle \phi _i^B\right| +\sigma _x\otimes \sum_ix_i^{\prime }p_i\left|
\phi _i^B\right\rangle \left\langle \phi _i^B\right| +\sigma _y\otimes
\sum_iy_i^{\prime }p_i\left| \phi _i^B\right\rangle \left\langle \phi
_i^B\right| +\sigma _z\otimes \sum_iz_i^{\prime }p_i\left| \phi
_i^B\right\rangle \left\langle \phi _i^B\right| \right] 
\end{array}
\right. 
\]
and 
\[
\left( 
\begin{array}{l}
x_i^{\prime } \\ 
y_i^{\prime } \\ 
z_i^{\prime }
\end{array}
\right) =R\left( 
\begin{array}{l}
x_i \\ 
y_i \\ 
z_i
\end{array}
\right) 
\]
The condition $I-R^TR\geq 0$ means that $\left| \overrightarrow{r_i^{\prime }%
}\right| ^2=x_i^{\prime 2}+y_i^{\prime 2}+z_i^{\prime 2}\leq 1$, therefore $%
\Upsilon _R\left( \rho _{AB}\right) $ is still a density matrix (positively
defined). This completes the proof of theorem 2.

And from eq. (\ref{sepms}), it's easy to get 
\[
M_0-\overrightarrow{r}\cdot \overrightarrow{M}=\frac 12\left[
\sum_ip_i\left( 1-\overrightarrow{r}\cdot \overrightarrow{r_i}\right) \left|
\phi _i^B\right\rangle \left\langle \phi _i^B\right| \right] 
\]
Part (i) of theorem 3 is obvious if we notice that $1-\overrightarrow{r}%
\cdot \overrightarrow{r_i}\geq 0$.

From eq. (\ref{sepms}), we also get 
\[
\begin{array}{lll}
tr\left( M_0^2-M_x^2-M_y^2-M_z^2\right)  & = & \frac 14\sum_{ij}p_ip_j\left(
1-x_ix_j-y_iy_j-z_iz_j\right) \left| \left\langle \phi _i^B\right. \left|
\phi _j^B\right\rangle \right| ^2 \\ 
& = & \frac 14\sum_{ij}p_ip_j\left( 1-\overrightarrow{r_i}\cdot 
\overrightarrow{r_j}\right) \left| \left\langle \phi _i^B\right. \left| \phi
_j^B\right\rangle \right| ^2 \\ 
& \geq  & 0
\end{array}
\]
In the last step the inequality $1-\overrightarrow{r_i}\cdot \overrightarrow{%
r_j}\geq 0$ has been used. This proves part (ii) of theorem 3.

For example, given 
\[
\rho_{AB} =\lambda \left| \phi^+ \right\rangle \left\langle \phi^+ \right|
+(1-\lambda) \left| \phi^- \right\rangle \left\langle \phi^- \right| 
\]
where $\left| \phi^{\pm} \right\rangle \equiv \frac{1}{\sqrt{2}} \left(
\left| 00 \right\rangle \pm \left| 11 \right\rangle \right)$. We know that
the above state is separable iff $\lambda=\frac{1}{2}$. From each of the
three theorems, we also get the same result.

But none of these theorems is a sufficient criterion of separability for
general mixed states.

Some other approaches are discussed in the following, they may be useful to
the search of a necessary and sufficient separability criterion.

Let a given state $\rho_{AB}$ be written in the spectrum representation as
\begin{equation}
\rho_{AB} = \sum_i \lambda_i \left| i^{AB} \right\rangle \left\langle i^{AB}
\right|
\end{equation}
where $\left| i^{AB} \right\rangle$ are the eigenstates of $\rho_{AB}$
corresponding to the eigenvalues $\lambda_i$. Let $\left| \Psi_{ABC}
\right\rangle$ define the purification of $\rho_{AB}$, 
\[
\left| \Psi_{ABC} \right\rangle \equiv \sum_i \sqrt{\lambda_i} \left| i^{AB}
\right\rangle \left| i^C \right\rangle 
\]
here $\{ \left| i^C \right\rangle \}$ is a set of orthonormal states of
system C. And 
\begin{eqnarray}
\rho_{AC} &\equiv &tr_B \left( \left| \Psi_{ABC} \right\rangle \left\langle
\Psi_{ABC} \right| \right)  \label{rhoac} \\
\rho_{BC} &\equiv & tr_A \left( \left| \Psi_{ABC} \right\rangle \left\langle
\Psi_{ABC} \right| \right)  \label{rhobc}
\end{eqnarray}
According to Hughston-Jozsa-Wootters' result \cite{hjw}, if $\rho_{AB}$ is
separable, then, there exists a unitary matrix $M$ such that 
\[
\sqrt{\lambda_i} \left| i^{AB} \right\rangle = \sum_j M_{ij} \sqrt{p_j}
\left| \psi_j^A \right\rangle \left| \phi_j^B \right\rangle 
\]
Expressed in terms of $\left| \psi_j^A \right\rangle$ and $\left| \phi_j^B
\right\rangle$, 
\begin{eqnarray}
tr_C (\rho_{AC}\rho_{BC}) &=& \sum_{jj\prime} p_j p_{j\prime} \left\langle
\psi_j^A | \psi_{j\prime}^A \right\rangle\left\langle \phi_{j\prime}^B |
\phi_{j}^B \right\rangle\left| \psi_j^A \right\rangle \left\langle
\psi_{j\prime}^A \right| \otimes\left| \phi_{j\prime}^B \right\rangle
\left\langle \phi_{j}^B \right| \\
\rho_{AB}^2 &=& \sum_{jj\prime} p_j p_{j\prime} \left\langle \psi_j^A |
\psi_{j\prime}^A \right\rangle\left\langle \phi_{j}^B | \phi_{j\prime}^B
\right\rangle\left| \psi_j^A \right\rangle \left\langle \psi_{j\prime}^A
\right| \otimes\left| \phi_{j}^B \right\rangle \left\langle \phi_{j\prime}^B
\right|
\end{eqnarray}
We find that these two expressions are very similar, the only difference is
that the second expression has all the index $j$ and $j\prime$ exchanged for
system B only. We get the above two expressions provided that $\rho_{AB}$ is
separable.

The relations between these two matrices $tr_C (\rho_{AC}\rho_{BC})$ and $%
\rho_{AB}^2$ may impose strong criterions for separability. But it is not
easy to find the exact relationship, a weaker criterion may be
found when we trace part of the system out.

Another possible approach is to investigate the properties of matrix
$N(\mu)$ which is defined by 
\[
N(\mu) \equiv (1+ \mu) \rho_A \otimes \rho_B - \mu \rho_{AB} 
\]
where $\rho_A$, $\rho_B$ are the reduced density matrix of $\rho_{AB}$. It
can be shown that $N(\mu)$ is also a state (i.e., semi-positive,
hermitian and traced to 1) when $\mu$ is around $0$.

In summary, we have presented three necessary separability criteria.
Our first separability criterion is very easy to apply, 
it involves only matrix multiplication, although it is not
as strong as the disorder criterion.
The other two criteria are presented
for bipartite systems of dimension $2\times N$; for such systems, Peres'
criterion can be obtained from our second criterion as a special case,
therefore it also helps us understand Peres' criterion.

We thank Stephen Fenner for useful discussion. This work was partially
supported by an NSF grant and an ONR grant.

\end{document}